\documentclass[letter,conference]{IEEEtran}
\IEEEoverridecommandlockouts
\usepackage{cite}
\usepackage{amsmath,amssymb,amsfonts}
\usepackage{algorithmic}
\usepackage{graphicx}
\usepackage{booktabs}
\usepackage{threeparttable}
\usepackage{stackengine}
\usepackage{textcomp}
\usepackage[aboveskip=5pt,belowskip=-10pt]{caption}
\usepackage{xcolor}

\newcommand{\xrowht}[2][0]{\addstackgap[.5\dimexpr#2\relax]{\vphantom{#1}}}
\def\BibTeX{{\rm B\kern-.05em{\sc i\kern-.025em b}\kern-.08em
    T\kern-.1667em\lower.7ex\hbox{E}\kern-.125emX}}
\renewcommand{\baselinestretch}{1.004}

\begin{document}

\title{Learned Intelligent Recognizer with Adaptively Customized RIS Phases in Communication Systems}

\author{\IEEEauthorblockN{Yixuan Huang${}^1$, Jie Yang${}^{2,3}$, Chao-Kai Wen${}^{4}$, Shuqiang Xia${}^{5,6}$, Xiao Li${}^{1}$, and Shi Jin${}^{1,3}$}
\IEEEauthorblockA{${}^1$National Mobile Communications Research Laboratory, Southeast University, Nanjing 210096, China \\
${}^2$Key Laboratory of Measurement and Control of Complex Systems of Engineering, Ministry of Education, \\
Southeast University, Nanjing 210096, China\\
${}^3$Frontiers Science Center for Mobile Information Communication and Security, Southeast University, Nanjing 210096, China\\
${}^4$Institute of Communications Engineering, National Sun Yat-sen University, Kaohsiung 80424, Taiwan\\
${}^5$ZTE Corporation, Shenzhen 518057, China\\
${}^6$State Key Laboratory of Mobile Network and Mobile Multimedia, Shenzhen 518057, China\\
Email: \{huangyx, yangjie, li\_xiao, jinshi\}@seu.edu.cn, chaokai.wen@mail.nsysu.edu.tw, xia.shuqiang@zte.com.cn
}
}

\newcommand{\bl}[1]{\textcolor{blue}{#1}}
\newcommand{\rl}[1]{\textcolor{red}{#1}}
\newcommand{\pl}[1]{\textcolor{purple}{#1}}

\maketitle

\begin{abstract}

This study presents an advanced wireless system that embeds target recognition within reconfigurable intelligent surface (RIS)-aided communication systems, powered by cuttingedge deep learning innovations.
Such a system faces the challenge of fine-tuning both the RIS phase shifts and neural network (NN) parameters, since they intricately interdepend on each other to accomplish the recognition task.
To address these challenges, we propose an intelligent recognizer that strategically harnesses every piece of prior action responses, thereby ingeniously multiplexing downlink signals to facilitate environment sensing.
Specifically, we design a novel NN based on the long short-term memory (LSTM) architecture and the physical channel model.
The NN iteratively captures and fuses information from previous measurements and adaptively customizes RIS configurations to acquire the most relevant information for the recognition task in subsequent moments.
Tailored dynamically, these configurations adapt to the scene, task, and target specifics.
Simulation results reveal that our proposed method significantly outperforms the state-of-the-art method, while resulting in minimal impacts on communication performance, even as sensing is performed simultaneously.

\end{abstract}

\section{Introduction}

Environment sensing is poised to be integrated into future wireless communication systems to enable ubiquitous sensing using channel state information (CSI), encompassing mapping, imaging, and recognizing \cite{huang2023joint}. Among these, target recognition has emerged as a pivotal issue for supporting ``context-aware'' applications. For instance, health monitoring and touchless human-computer interaction are facilitated through the recognition of human postures \cite{hu2020reconfigurable}, while classifying birds and drones enhances security surveillance \cite{costa2024static}.

Classification has been extensively explored in computer vision \cite{mnih2014recurrent}, inspiring some prior studies to design classifiers by first imaging the targets and then classifying their radio images \cite{saigre2022intelligent,huang2024ris}. However, radio imaging is inherently challenging, requiring extensive CSI measurements to capture detailed information about the targets, of which only a small portion is relevant to the recognition task \cite{saigre2022intelligent}. Thus, designing classifiers that directly map the limited measurements to class labels without imaging is considered more efficient. A hypothesis-testing-based method has been proposed in \cite{santos2024assessing}, but it incurs high complexity when calculating posterior probabilities across a large number of categories. In contrast, deep learning-based techniques employing fully connected (FC) networks have been utilized in \cite{zhao2023intelligent,hu2020reconfigurable} to mitigate this problem and significantly enhance classification accuracy.

Despite these advancements, the complex and unpredictable nature of wireless channels has inherently limited sensing accuracy. Recently, reconfigurable intelligent surface (RIS) technology has been leveraged to tailor electromagnetic environments for communication and sensing with low energy consumption \cite{sang2023multi}. Typically, random RIS phase shifts are used to gather diverse information about the target during sensing \cite{zhao2023intelligent}. Moreover, RIS configurations can be optimized by minimizing the averaged mutual coherence of the sensing matrix \cite{hu2020reconfigurable} or by training a principal component analysis-based dictionary \cite{liang2015reconfigurable}. Yet, these studies often optimize measurement acquisition and processing independently, neglecting to tailor RIS phases specifically for the classification task or fully utilize prior scene and task knowledge \cite{saigre2022intelligent}.

To address these challenges, a learned integrated sensing pipeline (LISP) is proposed in \cite{del2020learned}, integrating RIS phases as trainable physical variables within the neural network (NN). RIS phases and NN parameters are jointly optimized through supervised learning, generating RIS patterns specifically tailored for the scene and task, achieving state-of-the-art target recognition performance. Nevertheless, the LISP method \cite{del2020learned} optimizes all RIS configurations simultaneously, without considering that measurements with prior RIS patterns have been acquired before configuring the next phase shift.

In this study, we introduce the concept of jointly optimizing RIS phases and NN parameters, emphasizing that measurements obtained in previous moments contain information about the target, which can be leveraged to guide the design of RIS phases in subsequent moments. Inspired by \cite{mnih2014recurrent}, we employ a long short-term memory (LSTM) network to iteratively fuse acquired information and adaptively generate the tailored RIS pattern for the next moment, aiming at precise target recognition. By integrating physical information, our proposed method customizes both the hardware and software of the system, not only for the scene and task but also for the target being sensed, allowing for high classification accuracy with low measurement overhead. Additionally, the proposed sensing scheme can be conducted alongside communication processes, resulting in negligible impact on communication performance.

\section{System Model}
\begin{figure}
    \centering
    \includegraphics[width=0.69\linewidth]{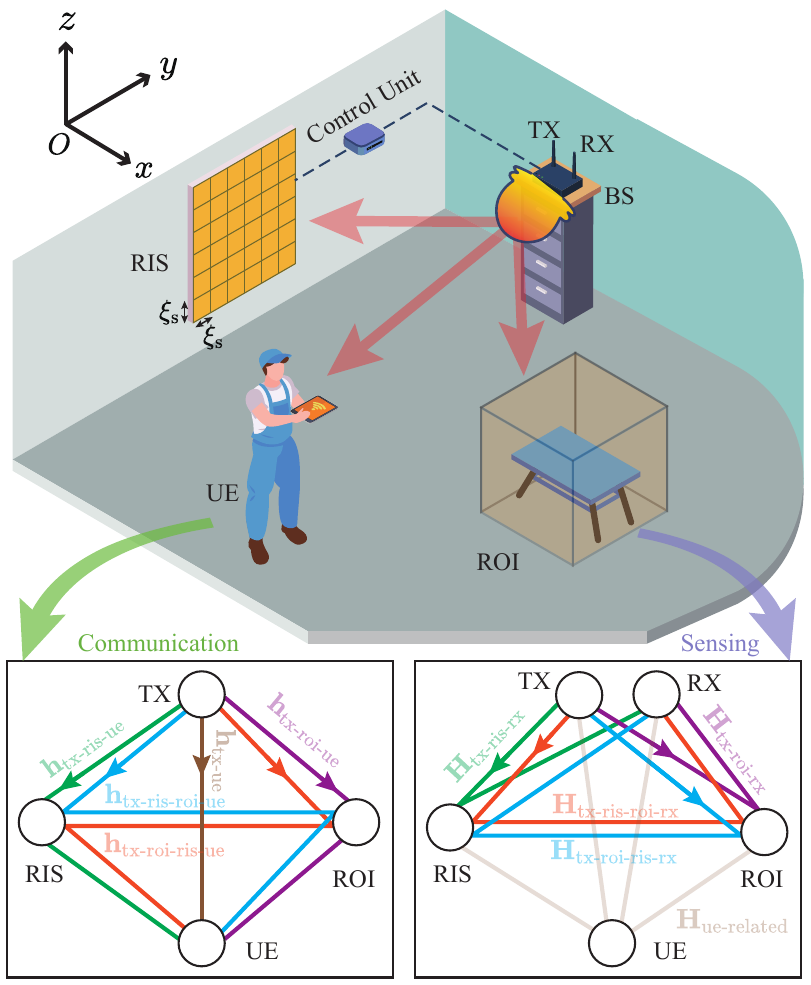}
    \captionsetup{font=footnotesize}
    \caption{Illustration of the proposed joint communication and sensing system with the aid of the RIS.}
    \label{fig-model}
\end{figure}

\addtolength{\topmargin}{0.05in}
We consider a RIS-aided communication system functioning within the 3D space $[x,y,z]^{\text{T}}\in\mathbb{R}^3$, as illustrated in Fig. \ref{fig-model}.
The full-duplex base station (BS) communicates with a single-antenna user equipment (UE) using orthogonal frequency division multiplexing (OFDM) signals.
We assume perfect self-interference cancellation between the BS transmitter (TX) and receiver (RX) through antenna separation \cite{zhang2015full}.
The TX and RX are uniform linear arrays configured with $N_{\text{t}}$ and $N_{\text{r}}$ antennas spaced at $\lambda/2$, respectively, where $\lambda$ is the wavelength.
The RIS comprises $N_{\text{s}}=N_{\text{y}}\times N_{\text{z}}$ elements, each of size $\xi_{\text{s}}\times\xi_{\text{s}}$.
Its phase shifts $\boldsymbol{\omega}\in\mathbb{C}^{N_{\text{s}}\times 1}$ are tuned by the BS to enhance communication and sensing performances.
The region of interest (ROI) can be discretized into $N_{\text{i}}$ voxels \cite{hu2020reconfigurable}, and its scattering coefficient image is represented by $\boldsymbol{\sigma}\in\mathbb{R}^{N_{\text{i}}\times 1}$.
The locations of the TX, RX, RIS, and ROI are known, with the distance between the RIS and the ROI denoted by $D$. 
The objective of our study is to identify the class of the target within the ROI during the communication process.

\subsection{Signal and Channel Models for Communication}

Consider a downlink (DL) communication scenario, where the TX transmits the signal $\mathbf{x}\in\mathbb{C}^{N_{\text{t}}\times 1}$ to the UE.
The received signal at the UE can be given as
\begin{equation}\label{eq-commun-signal}
{r}_{\text{com}} = \sqrt{P_{\text{t}}}\mathbf{h}_{\text{com}}^{\text{H}}\mathbf{x} + {n}_{\text{com}},
\end{equation}
where $\mathbf{h}_{\text{com}}^{\text{H}}\in\mathbb{C}^{1\times N_{\text{t}}}$ denotes the multipath channel from the TX to the UE, and ${n}_{\text{com}}\in\mathbb{C}$ is the additive Gaussian noise at the UE.
${P_{\text{t}}}$ presents the transmit power, and $\|\mathbf{x}\|_2 = 1$.
According to Fig. \ref{fig-model}, the channel $\mathbf{h}_{\text{com}}$ can be formulated as
\begin{equation}\label{eq-commun-channel}
\mathbf{h}_{\text{com}} =   \mathbf{h}_{\text{tx}\text{-}\text{ue}} + \mathbf{h}_{\text{tx}\text{-}\text{ris}\text{-}\text{ue}} + \mathbf{h}_{\text{tx}\text{-}\text{roi}\text{-}\text{ue}} 
  + \mathbf{h}_{\text{tx}\text{-}\text{ris}\text{-}\text{roi}\text{-}\text{ue}} + \mathbf{h}_{\text{tx}\text{-}\text{roi}\text{-}\text{ris}\text{-}\text{ue}},
\end{equation}
where $\mathbf{h}_{\text{tx}\text{-}\text{ue}}^{\text{H}}\in\mathbb{C}^{1\times N_{\text{t}}}$ denotes the line-of-sight (LOS) path from the TX to the UE.
$\mathbf{h}_{\text{tx}\text{-}\text{ris}\text{-}\text{ue}}^{\text{H}}$ and $\mathbf{h}_{\text{tx}\text{-}\text{roi}\text{-}\text{ue}}^{\text{H}}$ are the single-bounce paths scattered by the RIS and the target in the ROI, respectively.
$\mathbf{h}_{\text{tx}\text{-}\text{ris}\text{-}\text{roi}\text{-}\text{ue}}^{\text{H}}$ and $\mathbf{h}_{\text{tx}\text{-}\text{roi}\text{-}\text{ris}\text{-}\text{ue}}^{\text{H}}$ represent two twice-bounce paths.
The detailed forms of the cascaded channels can be found in Appendix \ref{appendix-channel}.
The multipaths that experience more bounces are assumed to be included in the noise ${n}_{\text{com}}$.

\subsection{Signal and Channel Models for Sensing}

The DL communication signal $\mathbf{x}$ can be simultaneously received by the RX after scattering of the RIS and the targets to realize environment sensing, given as
\begin{equation}
\mathbf{r}_{\text{sen}} = \sqrt{P_{\text{t}}}\overline{\mathbf{H}}_{\text{sen}}\mathbf{x} + \mathbf{n}_{\text{sen}},
\end{equation}
where $\mathbf{n}_{\text{sen}}$ is the additive noise at the RX.
$\overline{\mathbf{H}}_{\text{sen}}\in\mathbb{C}^{N_{\text{r}}\times N_{\text{t}}}$ denotes the multipath channel from the TX to RX, given as
\begin{equation}
\overline{\mathbf{H}}_{\text{sen}} = {\mathbf{H}}_{\text{ue}\text{-}\text{related}} + {\mathbf{H}}_{\text{sen}},
\end{equation}
where ${\mathbf{H}}_{\text{ue}\text{-}\text{related}}$ is the multipath related to the UE, and
\begin{equation}\label{eq-sensing-channel}
\begin{aligned}
{\mathbf{H}}_{\text{sen}} = & \ \mathbf{H}_{\text{tx}\text{-}\text{ris}\text{-}\text{rx}} + \mathbf{H}_{\text{tx}\text{-}\text{roi}\text{-}\text{rx}} + \mathbf{H}_{\text{tx}\text{-}\text{ris}\text{-}\text{roi}\text{-}\text{rx}} + \mathbf{H}_{\text{tx}\text{-}\text{roi}\text{-}\text{ris}\text{-}\text{rx}},
\end{aligned}
\end{equation}
denotes the CSI used for target recognition, where the direct path from the TX to the RX is assumed to have been perfectly removed.
The channels in \eqref{eq-sensing-channel} can be defined in similar forms to \eqref{eq-commun-channel}.
Since ${\mathbf{H}}_{\text{ue}\text{-}\text{related}}$ varies with the UE location and posture\footnote{
In scenarios like human posture recognition, the UE may be exactly the target in the ROI \cite{zhao2023intelligent}. Then, the channels related to the UE can be given as $\mathbf{H}_{\text{ue}\text{-}\text{related}}=\mathbf{H}_{\text{tx}\text{-}\text{roi}\text{-}\text{rx}} + \mathbf{H}_{\text{tx}\text{-}\text{ris}\text{-}\text{roi}\text{-}\text{rx}} + \mathbf{H}_{\text{tx}\text{-}\text{roi}\text{-}\text{ris}\text{-}\text{rx}}$. In this study, we consider a general model where the UE is not the target being sensed.},
we consider them additive disturbance to ${\mathbf{H}}_{\text{sen}}$.
The channel $\overline{\mathbf{H}}_{\text{sen}}$ can be estimated by the least squares (LS) algorithm \cite{yang2018beamspace} with the received signals of $N_{\text{t}}$ different DL signals, which are known at the BS\footnote{
The number of measurements for estimating ${\mathbf{H}}_{\text{sen}}$ may be reduced to be much lower than $N_{\text{t}}$ by harnessing the sparse property of ${\mathbf{H}}_{\text{sen}}$ \cite{yang2018beamspace}. In this study, we take the simple LS algorithm as an example for analysis.}.
Taking the channel estimation results as the measurement of ${\mathbf{H}}_{\text{sen}}$, we have
\begin{equation}\label{eq-sensing-measurement}
\widehat{\mathbf{H}}_{\text{sen}} = {\mathbf{H}}_{\text{sen}} + \mathbf{N}_{\text{sen}},
\end{equation}
where $\mathbf{N}_{\text{sen}}$ is the noise originated from $\mathbf{n}_{\text{sen}}$ and $\mathbf{H}_{\text{ue}\text{-}\text{related}}$.

\subsection{Protocol Design and Spectral Efficiency Analysis}

\begin{figure}
    \centering
    \includegraphics[width=0.88\linewidth]{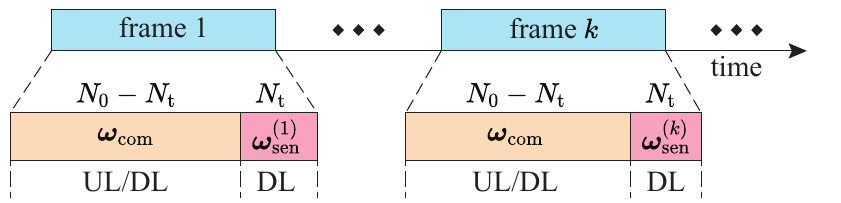}
    \captionsetup{font=footnotesize}
    \caption{The proposed protocol with time-division RIS configurations.}
    \label{fig-protocol}
\end{figure}

We design the protocol based on the 5G NR frame structure, considering its flexible uplink (UL)/DL switching \cite{lin20195g} and the fast RIS phase reconfiguration time \cite{sang2022coverage}.
We assume that the RIS assists in both communication and sensing.
To improve the communication performance, the RIS phases are optimized with the centralized algorithm proposed in \cite{wu2018intelligent}, given as $\boldsymbol{\omega}_{\text{com}}$, to maximize the spectral efficiency (SE).
The details of the optimization problem formulation can be found in Appendix \ref{appendix-optimization}.
To realize target recognition during communication, we propose that the TX should transmit DL signals, and the RIS phase shifts are configured according to the proposed method in Sec. \ref{sec-ris-phase-design} at the last $N_{\text{t}}$ symbol intervals in each frame, given as $\boldsymbol{\omega}_{\text{sen}}^{(k)}$, where $k=1, 2, \ldots, K$.
The $N_{\text{t}}$ symbols are simultaneously received by the UE and the RX to realize communication and sensing, respectively, where the RX derives the estimates of ${\mathbf{H}}_{\text{sen}}$ with the LS algorithm.
By varying the RIS phases with $K$ distinct configurations, the target label can be predicted.
The protocol is depicted in Fig. \ref{fig-protocol}, where $N_0 = 140\times 2^{\mu}$ symbolizes the number of OFDM symbols in one frame, and $\mu$ is the numerology in 5G NR.
The proposed protocol implements intermittent sensing intervals to match the delay of RIS phase generation in the proposed algorithm and to ensure high communication rates in each frame\footnote{The number of RIS phase changes in one frame may be increased to accelerate the sensing process, cooperating with the NN processing speed and potentially degrading the communication performance.}.

We assume that the system employs the comb-type pilot structure and estimate the DL communication channel $\mathbf{h}_{\text{com}}^{\text{H}}$ at each symbol interval.
Moreover, we assume that the locations of all the items in Fig. \ref{fig-model} are static in one frame.
Consequently, the channel $\mathbf{h}_{\text{com}}$ is only the function of $\boldsymbol{\omega}$, rewritten as $\mathbf{h}_{\text{com}}(\boldsymbol{\omega})$.
Denote the noise variance of ${n}_{\text{com}}$ as $\sigma^2_{\text{com}}$, the SE of DL communication with perfect CSI can be formulated as
\begin{equation}\label{eq-se1}
{\text{SE}}(\boldsymbol{\omega}) = \log_{2}\left(1+\frac{P_{\text{t}}\left\|\mathbf{h}_{\text{com}}(\boldsymbol{\omega})\right\|^{2}}{\sigma^2_{\text{com}}}\right).
\end{equation}
According to Fig. \ref{fig-protocol}, the average SE can be given as
\begin{equation}\label{eq-se2}
\overline{\text{SE}}_{\mu} = \frac{N_0-N_{\text{t}}}{N_0} {\text{SE}}(\boldsymbol{\omega}_{\text{com}}) + \frac{N_{\text{t}}}{N_0} {\text{SE}}(\boldsymbol{\omega}_{\text{sen}}).
\end{equation}
Since the antenna number $N_{\text{t}}$ is typically much smaller than $N_0$, the communication performance loss of the proposed protocol is considered tiny compared to ${\text{SE}}(\boldsymbol{\omega}_{\text{com}})$.

\begin{figure*}
    \centering
    \includegraphics[width=0.73\linewidth]{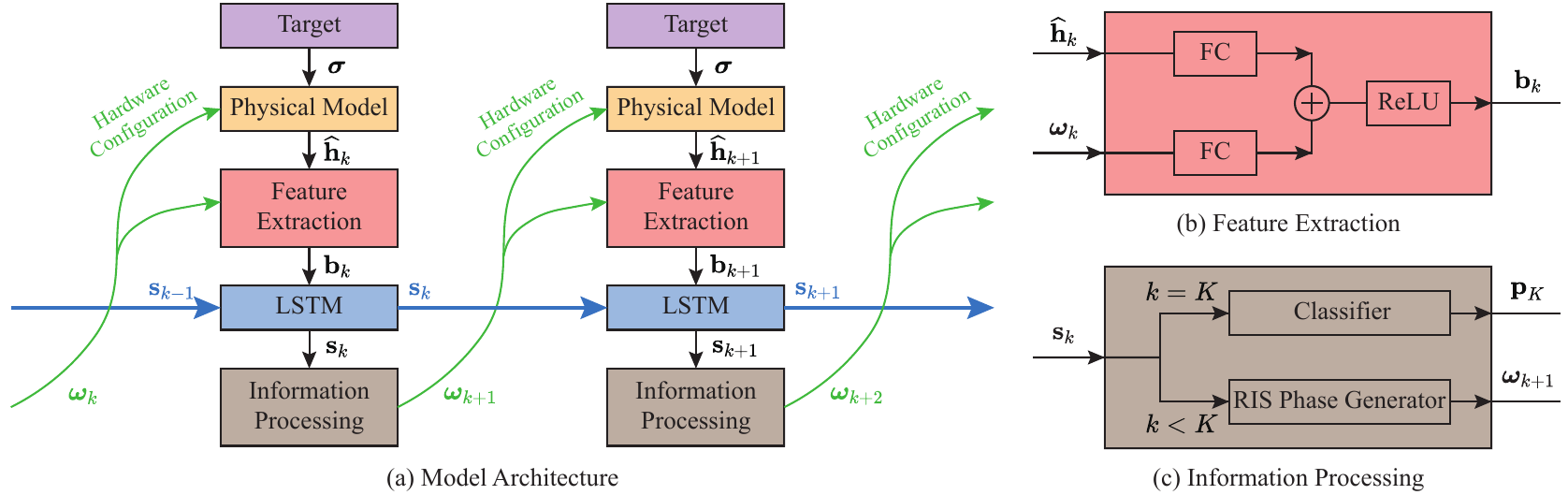}
    \captionsetup{font=footnotesize}
    \caption{The proposed NN architecture.}
    \label{fig-net}
\end{figure*}

\section{Learned Recognizer with Adaptive RIS Phase Customization}
\label{sec-ris-phase-design}

In this section, we focus on designing RIS phase shifts to enhance sensing accuracy while minimizing the number of RIS phase configurations, denoted as $K$. Given the highly coupled properties between the RIS patterns $\boldsymbol{\omega}_{\text{sen}}^{(k)}$ and the NN parameters $\boldsymbol{\theta}$, we propose to jointly learn their values through supervised learning. The RIS phase shifts are adaptively tailored to the scene, task, and target being sensed by harnessing the CSI ${\mathbf{H}}_{\text{sen}}$ obtained from previous configurations and integrating with the physical channel model.

\subsection{Overall Design of the Network}

Drawing on techniques from the field of computer vision \cite{mnih2014recurrent}, we base our NN on an LSTM architecture, as depicted in Fig. \ref{fig-net}(a). At each moment $k$, corresponding to the $k$-th frame in Fig. \ref{fig-protocol}, the proposed NN merges the information from the $k$ obtained measurements and adjusts the RIS configuration for the subsequent $(k+1)$-th moment to gather the most relevant information for identifying the target class. The newly generated RIS phase $\boldsymbol{\omega}_{\text{sen}}^{(k+1)}$ is then applied to the RIS hardware, acquiring a new measurement for further analysis. Our NN aims to simultaneously optimize the system's hardware and software components, in collaboration with the physical channel models.
The RIS phases are specifically tailored for each target at each moment, enabling the gradual recovery of the comprehensive information of the target's shape and scattering characteristics, thereby facilitating accurate recognition. Next, we detail the key modules of the proposed NN.

\subsection{Key Modules of the Network}

\textbf{Physical Model:}
This module is a reflection of the physical wave interactions, which projects the target image $\boldsymbol{\sigma}$ to the channel measurement $\widehat{\mathbf{h}}_k$ with the given RIS phase configuration $\boldsymbol{\omega}_k$.
Eliminating the subscript $(\cdot)_{\text{sen}}$, stacking the matrices into vectors, and considering the $K$ RIS configurations, \eqref{eq-sensing-measurement} can be rewritten as
\begin{equation}\label{eq-sensing-measurement-new}
\widehat{\mathbf{h}}_{k} = {\mathbf{h}}_{k} + \mathbf{n}_{k}, \quad k=1,2,\ldots,K.
\end{equation}
Under the assumption of static cascaded channels, the CSI ${\mathbf{h}}_{k}\in\mathbb{C}^{N_{\text{t}}N_{\text{r}}\times 1}$ is the function of the target scattering coefficient image $\boldsymbol{\sigma}$ and the RIS phase shift $\boldsymbol{\omega}_k$.
Thus, we have
\begin{equation}\label{eq-physical-model}
{\mathbf{h}}_{k} = f_{\text{phy}} (\boldsymbol{\sigma}, \boldsymbol{\omega}_{k}),
\end{equation}
where $f_{\text{phy}}$ corresponds to the physical channel model, whose detailed form is given in Appendix \ref{appendix-channel}.
$f_{\text{phy}}$ includes no learnable parameters, since the projection relationship shown in \eqref{eq-physical-model} is priorly known with the available locations of the items in Fig. \ref{fig-model}.

\textbf{Feature Extraction:}
This module extracts the information involved in $\widehat{\mathbf{h}}_k$ to a feature vector $\mathbf{b}_k\in \mathbb{R}^{B_1}$, where $B_1$ is the output dimension of the FC layers, as illustrated in Fig. \ref{fig-net}(b).
According to \cite{mnih2014recurrent}, we simultaneously input the RIS phase shift $\boldsymbol{\omega}_{k}$ and the measurement $\widehat{\mathbf{h}}_k$ to this module, guiding the NN to extract information about $\boldsymbol{\sigma}$.
This module can be formulated as
\begin{equation}
{\mathbf{b}}_{k} = f^{\boldsymbol{\theta}_{1}}_{\text{fea}} (\widehat{\mathbf{h}}_k, \boldsymbol{\omega}_{k}),
\end{equation}
where $\boldsymbol{\theta}_{1}$ is the learnable parameters.
Specifically, $\widehat{\mathbf{h}}_k$ and $\boldsymbol{\omega}_{k}$ are input to two independent FC layers, whose outputs are summed up and activated by the rectified linear unit (ReLU) function.
Since $\widehat{\mathbf{h}}_k$ and $\boldsymbol{\omega}_{k}$ are both complex vectors, we stack the real and imaginary parts of $\widehat{\mathbf{h}}_k$ to a real-value vector with the length of $2N_{\text{t}}N_{\text{r}}$, whereas only the phase information of $\boldsymbol{\omega}_{k}$ is reserved, whose elements have unit modulus.

\textbf{LSTM:}
This is the core module of the proposed NN.
It iteratively extracts and fuses the information lying in the feature vector $\mathbf{b}_k$ and the state vector $\mathbf{s}_{k-1}$, given as
\begin{equation}
{\mathbf{s}}_{k} = f^{\boldsymbol{\theta}_{2}}_{\text{lstm}} (\mathbf{b}_k, \mathbf{s}_{k-1}),
\end{equation}
where $\boldsymbol{\theta}_{2}$ is the learnable parameters.
${\mathbf{s}}_{k}\in\mathbb{R}^{B_2}$ denotes the state vector output by the LSTM module at the $k$-th moment, involving the target information lying in $[\widehat{\mathbf{h}}_{1}, \widehat{\mathbf{h}}_{2}, \ldots, \widehat{\mathbf{h}}_{k}]$.
The input and output dimensions of this module are $B_1$ and $B_2$, respectively.
With the accumulation of the measurements, rich information about the target is embedded into $\mathbf{s}_{k}$, which can be projected to one of the available categories.
Moreover, $\mathbf{s}_{k}$ may also reflect the absence of certain information, which is essential for the system to make a solid decision, guiding the RIS phase design.
The state vector $\mathbf{s}_{k}$ is transferred to the information processing module for class decision or hardware customization, as well as the LSTM module at the next moment for information fusion.

\textbf{Information Processing:}
This module takes in the state vector $\mathbf{s}_{k}$ and makes decisions with the extracted information contained in the acquired $k$ measurements.
Specifically, $\mathbf{s}_{k}$ is input to two sub-modules, the classifier and the RIS phase generator, as depicted in Fig. \ref{fig-net}(c).
The classifier only works when $k=K$ and outputs the probabilities that the target belongs to each category, denoted as $\mathbf{p}_K \in \mathbb{R}^{N_{\text{c}}\times 1}$, where $N_{\text{c}}$ is the number of possible categories.
The RIS phase generator works at each moment when $k<K$, adaptively generating the best RIS configuration $\boldsymbol{\omega}_{k+1}$ for the next moment, which is configured to the RIS hardware at the last $N_{\text{t}}$ symbol intervals of the $(k+1)$-th frame.
The two sub-modules are composed of two independent FC layers, given as
\begin{equation}
{\mathbf{p}}_{K} = f^{\boldsymbol{\theta}_{3}}_{\text{cla}} (\mathbf{s}_{K}),\quad {\boldsymbol{\omega}}_{k+1} = f^{\boldsymbol{\theta}_{4}}_{\text{pha}} (\mathbf{s}_{k}),
\end{equation}
where $\boldsymbol{\theta}_{3}$ and $\boldsymbol{\theta}_{4}$ are the learnable parameters of the classifier and the RIS phase generator, respectively.
Since the state vector $\mathbf{s}_{k}$ is unique for each target, the proposed NN generates different RIS phase shifts for distinct targets.
However, the first RIS pattern $\boldsymbol{\omega}_1$ is the same for each target, which is also learned through NN training by integrating it as part of the trainable parameters \cite{del2020learned}.

\subsection{Network Training}

In this study, we only consider continuous RIS phase shifts, thus, the NN $f_{\boldsymbol{\theta}}$ can be trained using the gradient descent algorithm.
For discrete RIS phase shifts, a temperature parameter can be introduced to realize back-propagation under quantization constraints \cite{saigre2022intelligent}.
The cross-entropy classification loss function is used in our study, given as
\begin{equation}\label{eq-cross-entropy}
L_{\text{CE}} = -\frac{1}{M}\sum_{m=1}^M\log(\mathbf{p}_{K}[c_m]),
\end{equation}
where $M$ is the number of training samples, and $c_m\in [1,N_{\text{c}}]$ is the index of the true target label for the $m$-th training data.
$\mathbf{p}_{K}[c_m]$ denotes the $c_m$-th element in vector $\mathbf{p}_{K}$.
$L_{\text{CE}}$ is minimized to optimize $\boldsymbol{\theta}=\{\boldsymbol{\theta}_1, \boldsymbol{\theta}_2, \boldsymbol{\theta}_3, \boldsymbol{\theta}_4, \boldsymbol{\omega}_1\}$, and the gradients are backpropagated through each of the modules.

\subsection{Comparison with Prior Studies}
The proposed NN showcases notable advancements over prior studies. It uniquely combines RIS configurations and NN parameters for joint optimization, diverging from the separated approach in \cite{zhao2023intelligent,hu2020reconfigurable,liang2015reconfigurable}.
Our NN customizes RIS phase shifts for individual targets and harnesses prior scene, task, and target information, enhancing system performance --- a strategy not previously explored.
Additionally, our NN adaptively generates RIS phase shifts, in contrast to the fixed trainable parameters in \cite{del2020learned}.
This adaptive capability leads to superior sensing performance. Despite its complexity and the time required to generate RIS configurations, our method employs a protocol with designed sensing intervals to maintain efficiency, as illustrated in Fig. \ref{fig-protocol}.

\section{Numerical Results}

\subsection{Simulation Settings}

We employ the MNIST dataset with $N_{\text{c}}=10$ to simulate the target in the ROI \cite{del2020learned}, where $M = 60,000$ training data and 10,000 testing data are used.
The pictures in the dataset are transformed into $30\times 30$ gray images, whose pixel values are subsequently normalized to $[0, 4\pi S^2/\lambda^2]$ \cite{huang2023joint}, representing the radar cross section of the voxel with the size of $\lambda\times\lambda$ in the 3D space, where $S$ is the voxel area.
The TX is located at $[30\lambda, 50\lambda, 50\lambda]^{\text{T}}$, and the antenna number $N_{\text{t}}=N_{\text{r}}=2$.
The RIS location is $[0, 0, 0]^{\text{T}}$ with the element size $\xi_{\text{s}} = \lambda/2$.
The ROI is centered at $[D, 0, 0]^{\text{T}}$, and the UE location is $[30\lambda, -50\lambda, 0]^{\text{T}}$.
The received noise power at the UE and the RX is set to -80 dBm.
For simplicity, we only employ the measurements on the center frequency for target sensing.

The NN training configurations include a batch size of 128, a total of 200 training epochs, and an initial learning rate of $10^{-3}$.
The validation set occupies 10\% of the training data.
The LSTM module contains a layer with $B_2 = 256$ hidden units, and its input size $B_1 = 256$.
All the FC layers consist of one hidden layer with 256 neurons.
The NN parameters are optimized with the Adam algorithm on an Nvidia 3090 GPU using the PyTorch platform.
We take the correct prediction rate $\eta$ as the performance evaluation metric.

\begin{figure}
    \centering
    \includegraphics[width=0.75\linewidth]{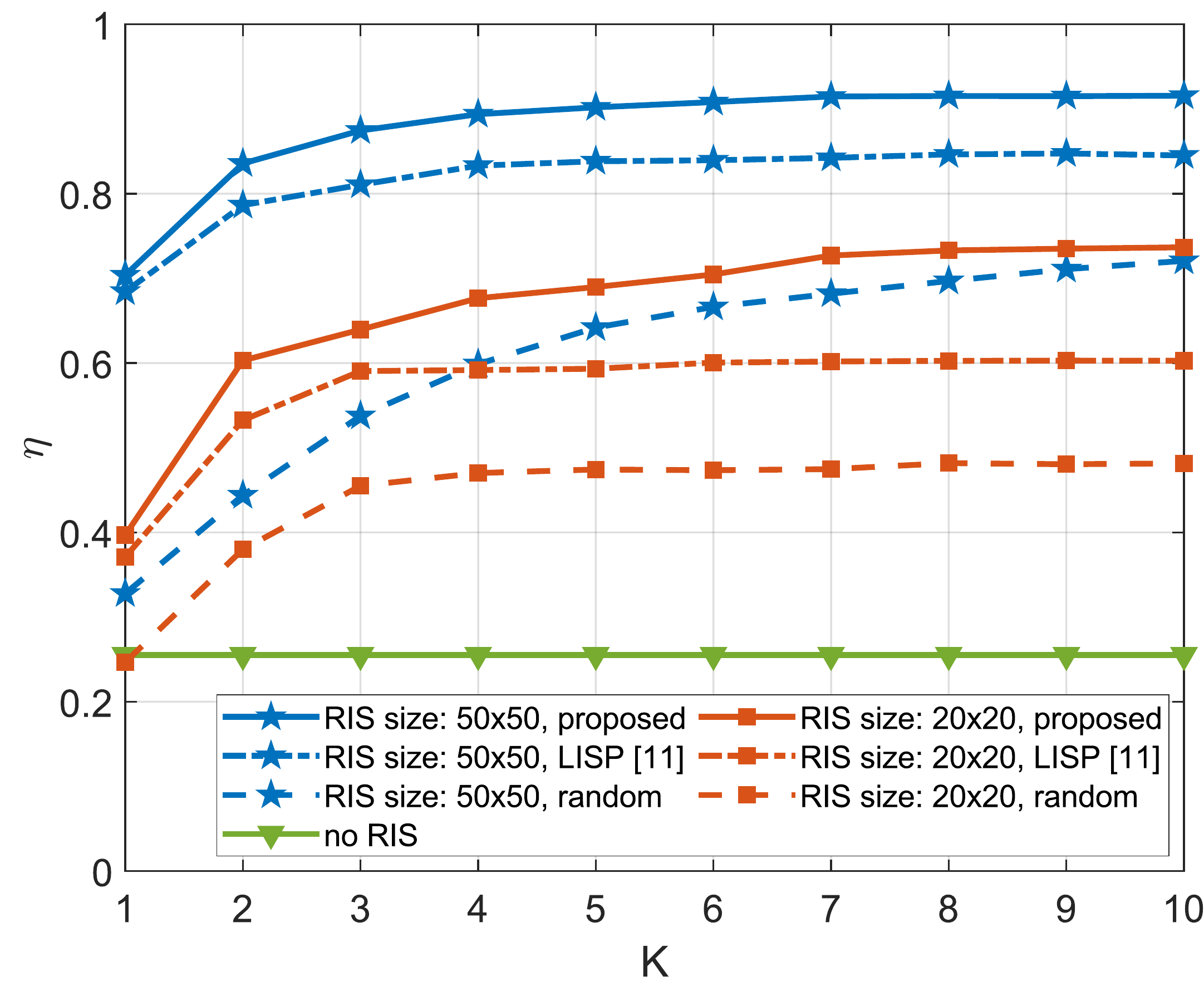}
    \captionsetup{font=footnotesize}
    \caption{Comparison of $\eta$ with different RIS array sizes and configurations.}
    \label{fig-result1-K}
\end{figure}

\subsection{Results and Discussions}

\subsubsection{Performance Comparison of Various RIS Phase Designs}
We evaluate the correct prediction rates, denoted as $\eta$, across different RIS phase designs using half of the training data set. Each method is subjected to the same training strategy. The simulation results, conducted at a distance of $D = 50\lambda$, are illustrated in Fig. \ref{fig-result1-K}. They reveal significant performance enhancements of our proposed NNs over the LISP method \cite{del2020learned} and random configurations. Notably, $\eta$ improves as the number of measurements increases, reaching a saturation point for our method when $K\ge 7$. In contrast, the $\eta$ for the LISP method remains relatively unchanged for $K\ge4$. The time required to recognize the target class in the considered scenario, where each frame lasts 10ms, is less than 0.1s. Enhancing the RIS array size can improve $\eta$; however, our method demonstrates more significant increases in $\eta$ with a smaller RIS. A $20\times20$ RIS employing $K=10$ of our proposed configurations achieves comparable sensing accuracy to a $50\times50$ RIS with random phase shifts, potentially offering savings on hardware costs. Comparing scenarios with and without an RIS underscores the benefits of incorporating the RIS to aid in target recognition.

\subsubsection{Influence of Distance, Transmit Power, and Training Data Size}
\begin{figure}
    \centering
    \includegraphics[width=0.83\linewidth]{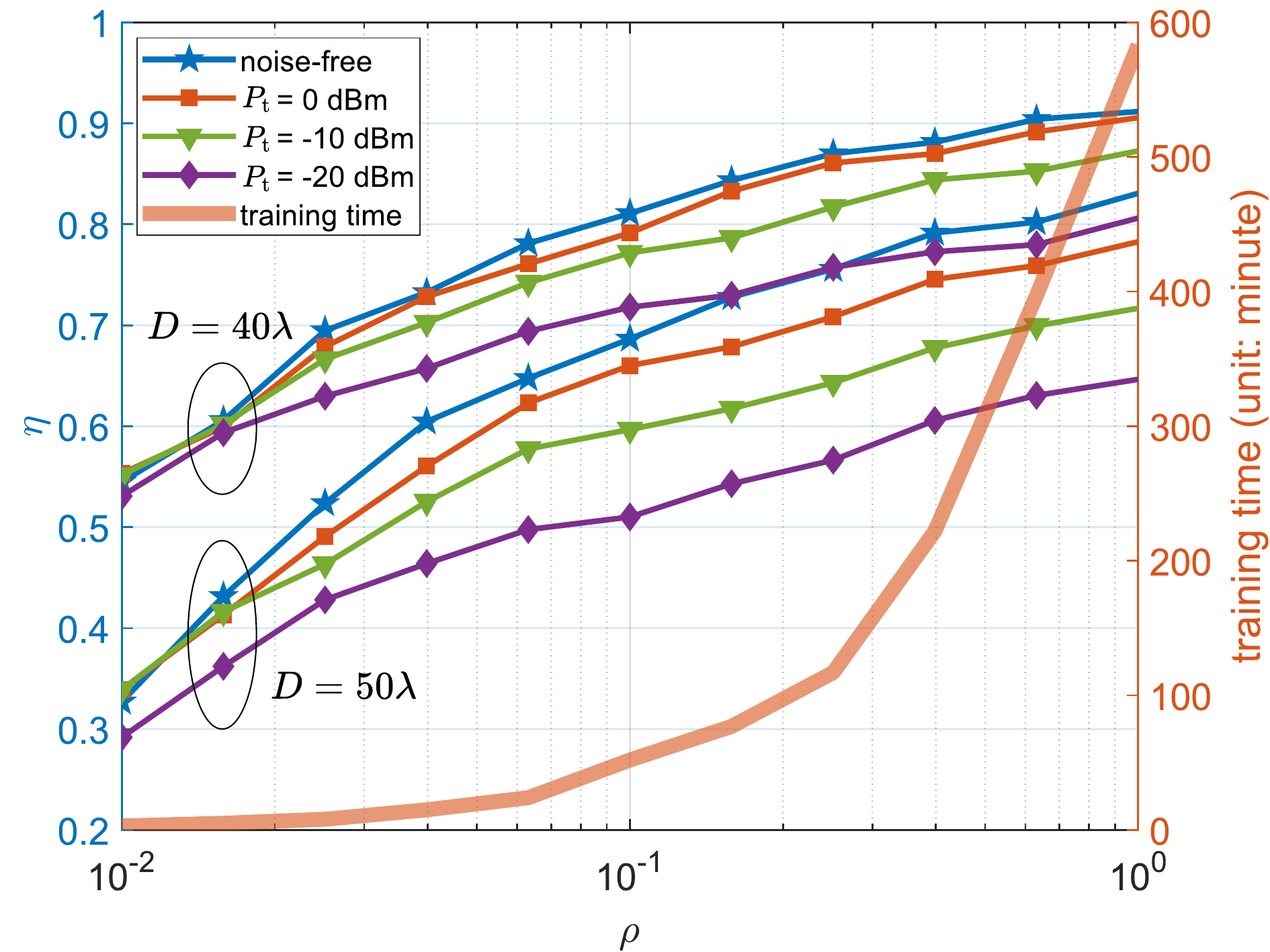}
    \captionsetup{font=footnotesize}
    \caption{$\eta$ with respect to $D$, $P_{\text{t}}$, and $\rho$.}
    \label{fig-result2-distance-noise}
\end{figure}

To assess the impact of target distance $D$, transmit power $P_{\text{t}}$, and the size of the training data, we utilize a $40\times40$ RIS. The additive noise at the RX is randomly generated during both the training and testing phases. The simulation results, presented in Fig. \ref{fig-result2-distance-noise}, show how $\rho$ affects the recognition performance, where $\rho$ is the ratio of the number of training samples used to the total available $M = 60,000$ instances. Training time was specifically noted for a noise-free environment at a distance of $D=40\lambda$. These results demonstrate that an increase in training data generally leads to higher sensing accuracy, albeit with an exponential growth in training time.
However, after completion of the training phase, the NN is capable of processing channel measurements and generating tailored RIS phases in less than 1 ms for each instance. Sensing performance declines with lower $P_{\text{t}}$. Moreover, a reduced distance $D$ significantly enhances recognition accuracy and mitigates the adverse effects of additive noise.

\begin{table}[t]
    \vspace{0.7cm}
    \renewcommand{\arraystretch}{1.3}
    \centering
    \fontsize{8}{8}\selectfont
    \captionsetup{font=small}
    \captionof{table}{SE performance comparison (unit: bit/s/Hz).}\label{tab-se}
    \begin{threeparttable}
        \begin{tabular}{ccccc}
            \specialrule{1pt}{0pt}{-1pt}\xrowht{10pt}
            RIS size & ${\text{SE}}(\boldsymbol{\omega}_{\text{com}})$ & ${\text{SE}}(\boldsymbol{\omega}_{\text{sen}})$ & $\overline{\text{SE}}_{\mu=1}$ & SE Loss \\
            \hline
            $10\times10$ & 14.94 & 13.39 & 14.93 & 0.07\% \\
            $20\times20$ & 17.20 & 13.58 & 17.17 & 0.17\% \\
            $30\times30$ & 19.10 & 13.50 & 19.06 & 0.21\% \\
            \specialrule{1pt}{0pt}{0pt}
        \end{tabular}
    \end{threeparttable}
    \vspace{-0.4cm}
\end{table}

\subsubsection{Communication Performance Analysis}
The SE performance, utilizing the proposed RIS configurations and protocol, is summarized in Table \ref{tab-se}, considering $\mu=1$ and $P_{\text{t}}=-10\ \text{dBm}$. When the phase shifts are optimized to enhance classification accuracy, SE experiences a minor reduction. This is because the LOS path between the TX and the UE primarily supports DL communication. Additionally, the phase shift $\boldsymbol{\omega}_{\text{sen}}$ is configured for only $N_{\text{t}}$ symbol intervals, significantly fewer than the total number of symbols $N_0$ in one frame. As a result, the average SE incurs only a marginal loss compared to ${\text{SE}}(\boldsymbol{\omega}_{\text{com}})$. Hence, our proposed approach achieves high sensing accuracy with minimal impact on communication performance.

\subsubsection{Correlation Analysis of Learned RIS Configurations}
\begin{figure}
    \centering
    \includegraphics[width=0.76\linewidth]{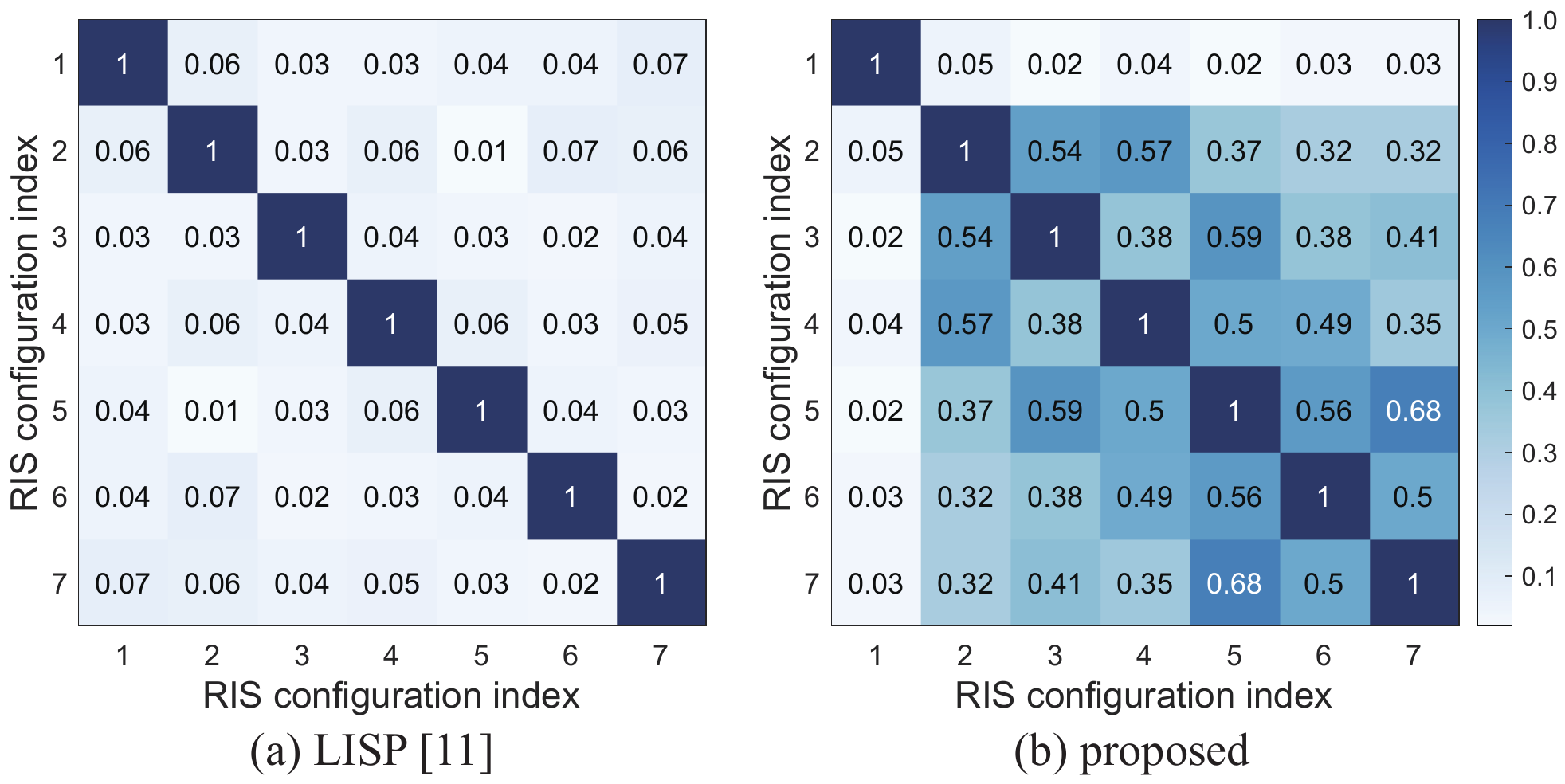}
    \captionsetup{font=footnotesize}
    \caption{Comparison of the correlations of the RIS phase configurations.}
    \label{fig-result3-heat}
\end{figure}

The proposed NN significantly diverges from the LISP method introduced in \cite{del2020learned}, particularly in the correlation patterns of the RIS phase shifts produced by both approaches. This comparison is depicted in Fig. \ref{fig-result3-heat} for $K=7$ and $N_{\text{s}} = 20\times 20$. Specifically, the RIS phase correlations are computed using the formula $|\boldsymbol{\omega}^{\text{H}}_{k_1}\boldsymbol{\omega}_{k_2}| / \|\boldsymbol{\omega}_{k_2}\|_2\|\boldsymbol{\omega}_{k_2}\|_2$, where $k_1, k_2 = 1, 2, \ldots, K$. In Fig. \ref{fig-result3-heat}(a), it is observed that the RIS configurations produced by the LISP method exhibit minimal correlations, thereby capturing pseudo-orthogonal information relevant to any target class. Conversely, as illustrated in Fig. \ref{fig-result3-heat}(b), the RIS patterns generated by the proposed NN show relatively high correlations from $k=2$ onwards. Given that the RIS phase shifts are uniquely tailored for each target, these results are averaged over 10,000 test data samples. Given our objective to capture the most pertinent information for classifying the target, where each target's category remains constant, it is logical to produce correlated RIS configurations. Thus, although the proposed method may not gather as much information as the LISP method, the information it does capture is specifically optimized for the targets, making it exceedingly valuable for the final classification task.

\section{Conclusion}
This study presents an intelligent recognizer with self-adaptive RIS configurations for communication systems, utilizing a novel LSTM-based neural network. This network adeptly integrates past measurement data to adaptively customize RIS phase shifts, optimizing both RIS and NN parameters based on prior scene, task, and target information. Simulations show our method outperforms existing algorithms with minimal impact on communication performance.

\section{Acknowledgement}
This work was supported in part by the Fundamental Research Funds for the Central Universities 2242022k60004, in part by the National Natural Science Foundation of China (NSFC) under Grants 62261160576, 62301156, 62341107 and 62231009, in part by the Key Technologies R\&D Program of Jiangsu (Prospective and Key Technologies for Industry) under Grants BE2023022 and BE2023022-1, in part by the Jiangsu Province Frontier Leading Technology Basic Research Project under Grant BK20212002, in part by the Fundamental Research Funds for the Central Universities 2242023K5003. The work of C.-K. Wen was supported in part by the National Science and Technology Council of Taiwan under the grant MOST 111-2221-E-110-020-MY3.

\begin{appendices}

\section{}\label{appendix-channel}
First, we give the channels in \eqref{eq-commun-channel} as $\mathbf{h}_{\text{tx}\text{-}\text{ue}} = \left(\frac{1}{(4\pi)^{0.5}d_{n_{\text{t}}, {\text{ue}}}} e^{-j2\pi \frac{d_{n_{\text{t}}, {\text{ue}}}}{\lambda}}\right)_{N_{\text{t}}\times 1}$,
% \begin{equation}\label{eq-appendix-a-1}
% \end{equation}
where $d_{n_{\text{t}}, {\text{ue}}}$ denote the distance between the UE and the $n_{\text{t}}$-th TX antenna.
Moreover,
$\mathbf{h}_{\text{tx}\text{-}\text{ris}\text{-}\text{ue}} = \mathbf{H}_{\text{ris}\text{-}\text{tx}} \text{diag}(\boldsymbol{\omega}) \mathbf{h}_{\text{ue}\text{-}\text{ris}}$,
$\mathbf{h}_{\text{tx}\text{-}\text{roi}\text{-}\text{ue}} = \mathbf{H}_{\text{roi}\text{-}\text{tx}} \text{diag}(\boldsymbol{\sigma}) \mathbf{h}_{\text{ue}\text{-}\text{roi}}$,
$\mathbf{h}_{\text{tx}\text{-}\text{ris}\text{-}\text{roi}\text{-}\text{ue}} = \mathbf{H}_{\text{roi}\text{-}\text{tx}} \text{diag}(\boldsymbol{\sigma}) \mathbf{H}_{\text{ris}\text{-}\text{roi}} \text{diag}(\boldsymbol{\omega}) \mathbf{h}_{\text{ue}\text{-}\text{ris}}$,
and
$\mathbf{h}_{\text{tx}\text{-}\text{roi}\text{-}\text{ris}\text{-}\text{ue}} = \mathbf{H}_{\text{ris}\text{-}\text{tx}} \text{diag}(\boldsymbol{\omega}) \mathbf{H}_{\text{roi}\text{-}\text{ris}} \text{diag}(\boldsymbol{\sigma}) \mathbf{h}_{\text{ue}\text{-}\text{roi}}$.
% The cascaded channels have the similar forms to \eqref{eq-appendix-a-1}.
Next, we give the details of $f_{\text{phy}}$ in \eqref{eq-physical-model}.
$f_{\text{phy}} (\boldsymbol{\sigma}, \boldsymbol{\omega}_{k}) = \text{vec}({\mathbf{H}}_{\text{sen}})$, where ${\mathbf{H}}_{\text{sen}}$ is defined in \eqref{eq-sensing-channel}, and $\text{vec}(\cdot)$ stacks a matrix into a vector.
We have
$\mathbf{H}_{\text{tx}\text{-}\text{ris}\text{-}\text{rx}} = \mathbf{H}_{\text{ris}\text{-}\text{rx}}\text{diag}(\boldsymbol{\omega})\mathbf{H}_{\text{tx}\text{-}\text{ris}}$,
$\mathbf{H}_{\text{tx}\text{-}\text{roi}\text{-}\text{rx}} = \mathbf{H}_{\text{roi}\text{-}\text{rx}}\text{diag}(\boldsymbol{\sigma})\mathbf{H}_{\text{tx}\text{-}\text{roi}}$,
$\mathbf{H}_{\text{tx}\text{-}\text{ris}\text{-}\text{roi}\text{-}\text{rx}} = \mathbf{H}_{\text{roi}\text{-}\text{rx}}\text{diag}(\boldsymbol{\sigma}) \mathbf{H}_{\text{ris}\text{-}\text{roi}} \text{diag}(\boldsymbol{\omega}) \mathbf{H}_{\text{tx}\text{-}\text{ris}}$,
and
$\mathbf{H}_{\text{tx}\text{-}\text{roi}\text{-}\text{ris}\text{-}\text{rx}} = \mathbf{H}_{\text{ris}\text{-}\text{rx}}\text{diag}(\boldsymbol{\omega}) \mathbf{H}_{\text{roi}\text{-}\text{ris}} \text{diag}(\boldsymbol{\sigma}) \mathbf{H}_{\text{tx}\text{-}\text{roi}}$.

\section{}\label{appendix-optimization}

In this section, we formulate the RIS phase optimization problem that maximizes the communication SE.
Denote $\mathbf{h}_{\rm{a}} = \mathbf{h}_{\text{tx}\text{-}\text{ue}} + \mathbf{h}_{\text{tx}\text{-}\text{roi}\text{-}\text{ue}}$, and $\mathbf{h}_{\rm{b}} = \mathbf{h}_{\text{tx}\text{-}\text{ris}\text{-}\text{ue}} + \mathbf{h}_{\text{tx}\text{-}\text{ris}\text{-}\text{roi}\text{-}\text{ue}} + \mathbf{h}_{\text{tx}\text{-}\text{roi}\text{-}\text{ris}\text{-}\text{ue}}$.
We have $\mathbf{h}_{\rm{b}} = \mathbf{H}_{\text{ris}\text{-}\text{tx}} \text{diag}(\boldsymbol{\omega}) (\mathbf{h}_{\text{ue}\text{-}\text{ris}} + \mathbf{H}_{\text{roi}\text{-}\text{ris}} \text{diag}(\boldsymbol{\sigma}) \mathbf{h}_{\text{ue}\text{-}\text{roi}}) + \mathbf{h}_{\text{tx}\text{-}\text{roi}\text{-}\text{ris}\text{-}\text{ue}}
= \mathbf{H}_{\text{ris}\text{-}\text{tx}} \text{diag}(\boldsymbol{\omega}) \mathbf{h}_{\rm{c}} + \mathbf{H}_{\rm{a}} \text{diag}(\mathbf{h}_{\text{ue}\text{-}\text{ris}}) \boldsymbol{\omega}
= \mathbf{H}_{\rm{b}}\boldsymbol{\omega}$,
where
$\mathbf{h}_{\rm{c}} = \mathbf{h}_{\text{ue}\text{-}\text{ris}} + \mathbf{H}_{\text{roi}\text{-}\text{ris}} \text{diag}(\boldsymbol{\sigma}) \mathbf{h}_{\text{ue}\text{-}\text{roi}}$,
$\mathbf{H}_{\rm{a}} = \mathbf{H}_{\text{roi}\text{-}\text{tx}} \text{diag}(\boldsymbol{\sigma}) \mathbf{H}_{\text{ris}\text{-}\text{roi}}$,
and
$\mathbf{H}_{\rm{b}} = \mathbf{H}_{\text{ris}\text{-}\text{tx}} \text{diag}(\mathbf{h}_3) + \mathbf{H}_{\rm{a}} \text{diag}(\mathbf{h}_{\text{ue}\text{-}\text{ris}})$.
To maximize the SE, we have to maximize $\left\|\mathbf{h}_{\text{com}}\right\|^{2} = \left\|\mathbf{H}_{\rm{b}}\boldsymbol{\omega} + \mathbf{h}_{\rm{a}}\right\|^{2}$, which possesses the same form as the objective function in \cite{wu2018intelligent}.

\end{appendices}

\bibliographystyle{IEEEtran}
\bibliography{ref}{}

\end{document}